\documentclass[11pt]{article}
\usepackage[english]{babel}
\usepackage{amsfonts}
\usepackage{amssymb}
\usepackage{amsmath}
\usepackage[dvips]{graphicx}
\usepackage{cite}
\usepackage{authblk}

\begin{document}

\title{LRS Bianchi I model with bulk viscosity in $f(R, T)$ gravity}


\author[1]{Siwaphiwe Jokweni\thanks{siwaphiwe14@gmail.com}}
\author[2]{Vijay Singh\thanks{gtrcosmo@gmail.com (corresponding author)}}
\author[3]{Aroonkumar Beesham\thanks{beeshama@unizulu.ac.za}}
\affil[1,2,3]{Department of Mathematical Sciences, University of Zululand, Private Bag X1001, Kwa-Dlangezwa, KwaZulu-Natal 3886, South Africa.}
\affil[3]{Faculty of Natural Sciences, Mangosuthu University of Technology,
Umlazi, South Africa.}

\date{}
\maketitle

\begin{abstract}

Locally-rotationally-symmetric Bianchi type-I viscous and non -viscous cosmological models are explored in general relativity (GR) and in $f(R, T)$ gravity. Solutions are obtained by assuming that the expansion scalar is proportional to the shear scalar which yields a constant value for the  deceleration parameter ($q=2$). Constraints are obtained by requiring the physical viability of the solutions. A comparison is made between the viscous and non-viscous models, and between the models in GR and in $f(R,T)$ gravity. The metric potentials remain the same in GR and in $f(R,T)$ gravity. Consequently, the geometrical behavior of the $f(R,T)$ gravity models remains the same as the models in GR. It is found that $f(R,T)$ gravity or bulk viscosity does not affect the behavior of  effective matter which acts as a stiff fluid in all models. The individual fluids have very rich behavior. In one of the viscous models, the matter either follows a semi-realistic EoS or exhibits a transition from stiff matter to phantom,  depending on the values of the parameter. In another model, the matter describes radiation, dust, quintessence, phantom, and the cosmological constant for different values of the parameter.  In general, $f(R,T)$ gravity diminishes the effect of bulk viscosity.

\end{abstract}



\section{Introduction}
\label{intro}

Our universe on a sufficiently large scale is homogeneous and isotropic. However, on smaller scales it is neither homogeneous nor isotropic. There are theoretical predictions that the early universe was also highly anisotropic which has been supported by many observations \cite{Netterfieldetal2002,Spergeletal2003,Andersonetal2013,BennettetalApJ2013,Hinshawetal2013,Adeetal2016,AghanimetalAA2018}. Among the simplest homogeneous and anisotropic  models,  Bianchi type-I (B-I) models play an outstanding role in understanding essential features of the early universe. Also in a universe filled with matter, the initial anisotropy in a B-I universe quickly dies away and the universe eventually becomes isotropic. Since the present-day universe is isotropic, the prominent features of the B-I models make them a prime candidate for studying the possible effects of anisotropy in the early evolution of the universe. In particular, the Locally-Rotationally-Symmetric (LRS) B-I spacetime is one of the simplified versions of the B-I model. In light of its importance, many researchers have  studied the LRS B-I models in various contexts (see \cite{BelinchonASS2006,SinghBeeshamGRG2019,SinghBeeshamIJMPD2019,SinghBeeshamEPJP2020,SinghBeeshamApSS2020} and references therein).

Though the present-day universe undergoes an accelerated expansionary evolution and bulk viscosity plays a very vital role in explaining this phenomenon. However, it does not exclude the existence of a decelerating phase in the early history of our universe. Mak and Harko \cite{MakHarkoIJMPD2002}, studied a causal bulk viscous cosmological fluid for a flat constantly decelerating B-I spacetime model, and showed that the model leads to a self-consistent thermodynamic description which  could describe a well-determined period of the evolution of our universe. Therefore, decelerating models have their own importance to understand the early evolution of the universe.

On the other hand, although a perfect fluid satisfactorily accounts for the large scale matter distribution in the universe, the realistic cosmological scenario requires the consideration of matter other than a perfect fluid. Some observed physical phenomena such as the large entropy per baryon and the noteworthy degree of isotropy of the cosmic background radiation, suggest dissipative effects in cosmology. Entropy producing processes and dissipative effects play a very significant role in the early evolution of the universe. In fluid cosmology, the simplest phenomenon associated with a non-vanishing entropy production is  bulk viscosity (for more detail see the review article \cite{GronASS1990} and references therein).

There are several processes which generates viscous effects (see Ref. \cite{SinghBeeshamAJP1999} for a list of some principal processes). The presence of bulk viscosity inaugurates many interesting features in the dynamics of the universe. Initially, it was proposed that neutrino viscosity could smooth out initial anisotropies and result the isotropic universe that we see today. The presence of bulk viscosity can avert the big-bang singularity too. Bulk viscosity can also explain a phenomenological process of particle creation in a strong gravitational field. The back-reaction effects of string creation can be modeled by a bulk viscous fluid. It has attracted much interest across the field of cosmology and many investigators have  pondered the effects of bulk viscosity in different contexts (see for examples \cite{PavonetalCQG1991,BeeshamPRD1993,TriginerPavonGRG1994,ZimdahlPRD1996,BeeshamNCB1996,ArbabGRG1997,ChimentoetalCQG1997,GavrilovetalCQG1997,SinghBeeshamGRG2000,ZimdahletalPRD2001,CataldoetalPLB2005,FabrisetalGRG2006,SzydlowskiHrycynaAP2007,SinghetalCQG2007,ColisteteetalPRD2007} and references therein). Most of these investigations are based on isotropic cosmology. However, in the search for a realistic picture of the early universe, a large number of studies have been done in anisotropic spacetimes as well \cite{RoyPrakashIJPAM1977,BanerjeeSantosGRG1984,BaliJainASS1991,BurdColeyCQG1994,HoogenColeyCQG1995,BaliJainASS1997,PradhanSinghIJMPD2004,sahaMPLA2005,SinghChaubeyPJP2007,BaliKumawatPLB2008,ChakrabortyRoyASS2008,SinghBaghelIJTP2010,RamVermaASS2010,BaghelSinghRAA2012,RametalJMP2012,SinghetalIJP2013}. The general B-I spacetime models also have been studied by many authors \cite{BurdColeyCQG1994,ChakrabortyRoyASS2008,BanerjeeSantosJMP1983,BanerjeeetalJMP1985,GoennerKowalewskiGRG1989,RomanoPavonPRD1993,ArbabGRG1998,BeeshametalGRG2000,BelinchonASS2005,SahaRikhvitskyPD2006,SinghKumarIJTP2009}. More specifically, some authors \cite{HuangJMP1990,PradhanOtarodASS2007,WangCPL2003} presented LRS B-I bulk viscous cosmological models.

On the other hand, the shortcomings of the $\Lambda$CDM model has confronted many authors to seek various alternatives to the fundamental theories of cosmology and astrophysics, which include modifications of general relativity (GR) itself by imposing extra terms in the Einstein-Hilbert action. The modified theories of gravity include higher derivative theories, Gauss-Bonnet $f(G)$ gravity, $f(R)$ theory,  $f(T)$ and $f(R,T)$ gravity theories. In the past decade,  $f(R,T)$ gravity has attracted the attention of many researchers to look at  many astrophysical and cosmological phenomena in the context of this theory (see \cite{SinghBeeshamEPJP2020} for a broad list of references).

Mahanta \cite{MahantaASS2014} considered a bulk viscous LRS B-I model in $f(R,T)$ gravity. The author assumed an  expansion scalar proportional to the shear scalar to solve the field equations. However, due to wrong signs considered in the field equations, his solutions are mathematically and physically invalid. Soon after, Shamir \cite{ShamirEPJC2015} presented solutions for the same model but without bulk viscosity. Later on, Sahoo and Reddy \cite{SahooReddyAp2018} considered an LRS B-I model containing bulk viscous matter in $f(R,T)$ gravity using a different deceleration parameter. Very recently, Yadav et al. \cite{YadavetalMPLA2019} have discussed the general B-I bulk viscous model in $f(R,T)=R+\lambda RT$ gravity with a hybrid expansion law of the scale factor.

Our purpose in this paper is to reconsider the model formulated by Mahanta \cite{MahantaASS2014} with correct field equations. We eloquently explore the behavior of the model keeping in view the physical viability of the model. Before considering the $f(R,T)$ gravity model, we first discuss the solutions in GR in the presence and absence of bulk viscosity. In this way, we distinguish the outcomes of the $f(R,T)$ gravity model with that of GR and recognize the role of $f(R,T)$ gravity and bulk viscosity.

Also, Mahanta \cite{MahantaASS2014} in $f(R,T)=R+2\lambda T$ model merely found the expression for the coefficient of bulk viscosity. While, in $f(R,T)=R+2\lambda T^2$ model, the author also studied the behavior of matter by considering two different forms of the  bulk viscosity coefficient. We implement this approach to the $f(R,T)=R+2\lambda T$ model. Therefore, our solutions are also an extension of Mahanta's work. It is worthwhile to mention that though a single matter content is considered in $f(R,T)$ gravity, due to the coupling between the trace, $T$ and the matter, some extra terms appear in the field equations. We treat these additional terms as coupled matter. We study the nature of this additional matter and its contribution to the cosmic evolution.

The work is organized as follows. An LRS B-I spacetime model in the presence and absence of bulk viscosity within the framework of GR is studied in Sect. 2 and in its subsections. The $f(R, T)=R+2\lambda T$ gravity viscous and non-viscous models are explored in Sect. 3 and in its subsections. The findings are accumulated in the concluding Sect. 4.

\section{The model in Einstein's gravity}

The spatially homogeneous and anisotropic LRS B-I space-time metric is given as
\begin{equation}
ds^{2} =dt^{2}-A^2dx^2-B^2(dy^2+dz^2),
\end{equation}
where $A$ and $B$ are the scale factors, and are functions of the cosmic time $t$.

The average scale factor and average Hubble parameter, respectively, are defined as
\begin{eqnarray}
a&=&(AB^2)^{\frac{1}{3}},\\
H&=&\frac{1}{3}\left(\frac{\dot A}{A}+2\frac{\dot B}{B}\right).
\end{eqnarray}
where a dot represents a derivative with respect to $t$.
We consider the energy-momentum tensor of the matter as
\begin{equation}
T_{ij}=(\rho+p)u_i u_j -pg_{ij},
\end{equation}
where $\rho$ is the energy density and $p$ is the thermodynamic pressure of the matter.  In comoving coordinates, $u^i = \delta^i_0$, where $u_i$ is the four-velocity of the fluid that satisfies the condition $u_i u^j =1$.

The Einstein field equations are given by
\begin{equation}
R_{ij}-\frac{1}{2}R\textsl{g}_{ij} =T_{ij},
\end{equation}
where $8\pi G=1=c$ are assumed.
The field equations (5) for the metric (1), with the consideration of the energy-momentum tensor (4), yield
\begin{eqnarray}
\left(\frac{\dot B}{B}\right)^2+2\frac{\dot A\dot B}{A B}&=&\rho,\\
\left(\frac{\dot B}{B}\right)^2+2\frac{\ddot B}{ B}&=&-p,\\
\frac{\ddot A}{A}+\frac{\ddot B}{B}+\frac{\dot A \dot B}{AB}&=&-p.
\end{eqnarray}
These equations consist of four unknowns, namely, $A$, $B$, $p$, $\rho$. Therefore, in order to find exact solutions, one supplementary constraint is required.

Mahanta \cite{MahantaASS2014} considered the expansion scalar, $\theta (=3H)$ to be  proportional to the shear scalar\footnote{$\sigma^2=\frac{1}{3}\left(\frac{\dot A}{A}-\frac{\dot B}{B}\right)^2$}, $\sigma$, which leads to
\begin{equation}
A=B^n,
\end{equation}
where $n$ is an arbitrary constant.
From (7) and (8), by the use of (9), one gets
\begin{equation}
\frac{\ddot{B}}{B}+(n+1)\left(\frac{\dot B}{B}\right)^2=0,
\end{equation}
which gives
\begin{equation}
B=\beta\left[(n+2)t+c_2\right]^{\frac{1}{n+2}}.
\end{equation}
Consequently
\begin{equation}
A=\alpha\left[(n+2)t+c_2\right]^{\frac{n}{n+2}}.
\end{equation}
The energy density and pressure become equal
\begin{equation}
\rho=p=\frac{(1+2n)}{(2+n)^2 t^2}.
\end{equation}
Hence, the effective matter behaves as stiff fluid. The energy density must be positive for a realistic cosmological scenario which is possible only for $n>-1/2$.

In section ``3" of the paper, Mahanta \cite{MahantaASS2014} worked out some geometrical parameters, namely, the volume, expansion scalar and shear scalar. All these parameters are defined in terms of the metric potentials $A$ and $B$. We see that the scale factors given in Eqs. (11) and (12) are identical to those of Mahanta's work though we have obtained these metric potentials in GR. In fact, left hand side (LHS) of the field equations in GR and in $f(R,T)$ gravity remains the same, only the right hand side (RHS) is different. When one simplifies equations (7) and (8) or ``(19)" and ``(20)" in Mahanta's paper, the RHS is cancelled out irrespective of the theory or even whatever may be the matter content (viscous or non-viscous) filled in the model. Hence, the metric potentials are independent of $f(R,T)$ gravity and matter content considered in such formulation. Consequently, all the geometrical parameters remain independent from $f(R,T)$ gravity and from matter content. Thus, the geometrical behavior of the model remains identical to the model in GR. We refer to Ref. \cite{ShamirEPJC2015} for the details of geometrical behavior of the model.

\subsection{Viscous model}

The energy density of bulk viscous matter remains the same but the pressure in energy-momentum tensor (4) for viscous fluid modifies as
\begin{equation}
\bar p=p'_m-\xi\theta,
\end{equation}
where $p'_m$ is the pressure of matter and $\xi$ is the coefficient of bulk viscosity.

The field equations for a viscous model remain almost similar to (6)--(8) except that the pressure $p$ is replaced by bulk viscous pressure $\bar p$. Therefore, the assumption (9) again leads to the solution (13), i.e., $\rho=\bar p$ which is identical to the non-viscous model. Hence, the bulk viscosity does not affect the behavior of effective matter and it acts as stiff matter. However, it is to be noted that the new field equations consist five unknowns, namely, $A$, $B$, $\rho$, $p'_m$, and $\xi$. Therefore, to determine the exact solutions completely, we require one more constraint other than (9). We have two ways: first, assuming an EoS that relates $\rho$ to $p'_m$, and then determine $\xi$; and second, assuming an explicit form for $\xi$ and then determine $\bar p$. We shall follow both approaches in the following section.

\subsubsection{The behavior of bulk viscous coefficient}

We assume that the matter follows the perfect fluid EoS
\begin{equation}
p'_m=\omega \rho,
\end{equation}
where $0\leq \omega\leq 1$ is the EoS parameter.

From (14), the expression for the coefficient of bulk viscosity is obtained as
\begin{equation}
\xi(t) = \frac{(2n+1)(\omega-1)}{\left(n+2\right)^2 t}.
\end{equation}
Since we have $n>-1/2$ for the energy density to be positive, the coefficient of bulk viscosity for any kind of matter except stiff matter ($\omega=1$) remains negative and increases with the evolution of the universe, for example, ultra-relativistic radiation ($\omega=1/3$), non-relativistic dust ($\omega=0$) or even for vacuum energy ($\omega=-1$). Also, as $\xi\to0$ when $t\to\infty$, the effect of bulk viscosity disappears at late times. In case of stiff matter, the coefficient of bulk viscosity vanishes and the solutions obtained in (13) are recovered.

\subsubsection{The behavior of matter}

By assuming a perfect fluid EoS, in Sects. ``3" and ``4.1", Mahanta \cite{MahantaASS2014} merely obtained the expression for the coefficient of bulk viscosity. However, in Sect. ``4.2" while considering $f(R,T)=\lambda R+\lambda T^2$, the author also considered two different relations between the bulk viscous coefficient and expansion scalar to study the properties of  matter and viscous fluid. However, other than the wrong signs in the field equations, there is another flaw in the model of $f(R,T)=\lambda R+\lambda T^2$. The author over-determined the solutions in this model. One needs two constraints to close the system  but the author used three, i.e., ``(21)", ``(61)" and the perfect fluid EoS, i.e., $p=\epsilon \rho$, $0\leq\epsilon\leq1$. Regardless of over determining the solutions, the sign on the right hand side of the field equations is also incorrect. Though we are not incorporating this model in the present study, but we shall use two assumptions those considered by Mahanta \cite{MahantaASS2014} in his model $f(R,T)=\lambda R+\lambda T^2$. These assumptions are: (i) the coefficient of bulk viscosity is inversely proportional to  the expansion scalar, i.e., $\xi \theta= k$, where $k$ is a positive constant; and (ii) the product of bulk viscosity coefficient and expansion scalar is directly proportional to energy density, i.e., $\xi \theta= k_1 \rho$, where $k_1>0$ is a constant. We consider both in in following cases to examine the nature of matter.

\subsubsection*{Case (i) $\xi \theta= k$}

In this case, the EoS parameter, $\omega'=p'_m/\rho$ gives
\begin{equation}
\omega'=1+\frac{k(2+n)^2 t^2}{1+2n}.
\end{equation}
At the origin, we have $\omega'=1$ (stiff matter). Mahanta considered only the case when $k>0$. If $k>0$, the EoS parameter starts from $\omega'=1$ and increases with the evolution. This case corresponds to a semi-realistic EoS $p = \varepsilon p~ (\varepsilon\geq 1)$. Many researchers \cite{BuchdahlLandAMS1968,WhittakerPRSL1968,IbanezSanzJMP1982} have studied cosmological models with the semi-realistic matter in forward approaches. However, if $k<0$, the EoS parameter renders interesting behavior. It exhibits a smooth transition from $\omega'=1$ (stiff matter) to $\omega'\to-\infty$ (phantom matter). Thus, it describes all kinds of known matter (stiff matter, radiation and dust) including the hypothetical form of dark energy (quintessence and phantom) and cosmological constant as well. Though the model only describes the decelerated universe, the dark energy characteristics anyway do not contradict because the matter which is showing this characteristic is not the effective matter in this model. We have already seen that the effective matter behaves as a stiff fluid.

\subsubsection*{Case (ii)  $\xi \theta= k_1 \rho$}

The EoS parameter in this case takes a constant value
\begin{equation}
\omega'=1+k_1.
\end{equation}
Hence, if $k_1>0$, the matter in this case also follows the semi-realistic EoS. On the other hand, if $k_1<0$, the model renders a variety of matter depending on the values of $k_1$, e.g., $\omega'=1/3$ (radiation) for $k_1=-2/3$, $\omega'=0$ (dust) for $k_1=-1$, $\omega'=-1/3$ (quintessence) for $k_1=-4/3$, $\omega'=-1$ (cosmological constant) for $k_1=-2$, and $\omega'<-1$ (phantom) when $k_1<-1$. If $k_1=0$, we have $\omega'=1$ (stiff matter), which implies $\xi=0$ as $\theta=1/t\neq0$. Hence, in the  absence of bulk viscosity, the solutions obtained in (13) are recovered.

\section{The model in $f(R, T) $ gravity}

It is vital to note that  $\rho$ and $p$ in Sect. 2. are the effective energy density and pressure, respectively, while in $f(R,T)$ gravity both the physical qualities no longer epitomize the effective energy density and pressure. The coupling between geometry and matter in $f(R,T)$ gravity adds some additional terms visible on the RHS of the field equations. These terms must be treated as matter that can be called coupled matter. Therefore, to distinguish between the main matter and coupled matter, we replace $p$ with $p_m$ and $\rho$ with $\rho_m$, which represent the primary or main matter. The notations for the energy density and pressure of the coupled matter are defined in Sect. 3.1.

The field equations in $f(R, T)= R+2f(T)$ gravity with the system of units $8\pi G=1=c$, are obtained as
\begin{equation}
R_{ij}-\frac{1}{2}R \textsl{g}_{ij} = T_{ij}+2\left(T_{ij}+p_m \textsl{g}_{ij}\right)f^\prime(T)+f(T) \textsl{g}_{ij},
\end{equation}
where a prime stands for a derivative with respect to the trace, $T$. For $f(T)=\lambda T$, i.e., $f(R, T)= R+2\lambda T$, where $T=g^{ij}T_{ij}=\rho_{m}-3p_{m}$, (19) simplifies as
\begin{equation}
R_{ij}-\frac{1}{2}R \textsl{g}_{ij}=(1+2\lambda)T_{ij}+\lambda(\rho_{m}-p_m)\textsl{g}_{ij},
\end{equation}
which for the metric (1) and the energy-momentum tensor (4), yield
\begin{eqnarray}
\left(\frac{\dot B}{B}\right)^2+2\frac{\dot A\dot B}{A B}&=&(1+3\lambda)\rho_m-\lambda p_m,\\
\left(\frac{\dot B}{B}\right)^2+2\frac{\ddot B}{ B}&=&-(1+3\lambda)p_m +\lambda\rho_m,\\
\frac{\ddot A}{A}+\frac{\ddot B}{B}+\frac{\dot A \dot B}{AB}&=&-(1+3\lambda)p_m +\lambda\rho_m.
\end{eqnarray}
This is the correct set of the field equations. One can see that the terms on the RHS of these equations are different from Eqs. ``(18)--(20)" in Ref. \cite{MahantaASS2014}.

Using (11) and (12) in (21) and (22) provided $\lambda\ne -1/2$, we obtain
\begin{equation}
\rho_{m}=p_{m} =\frac{(2n+1)}{(1+2\lambda)(n+2)^2 t^2}.
\end{equation}
This is the correct expression for the energy density and pressure which is different from incorrect one obtained in Eq. ``(26)" by Mahanta \cite{MahantaASS2014}. The primary matter acts as stiff matter. The energy density and pressure decrease with the evolution. The energy density ought to be positive for any physical viable cosmological model which is possible either
\begin{eqnarray}\nonumber
\lambda>-1/2\;\; &\text{if}& \;\;n>-1/2, \\
\text{or}\;\;\lambda<-1/2\;\; &\text{if}& \;\; n<-1/2.
\end{eqnarray}
Thus, while the solutions in GR are valid only for $n>-1/2$,  $f(R,T)$ gravity makes them valid for $n<-1/2$ also.

It is worthwhile to mention here that we have obtained  expression (24) without bulk viscosity but Mahanta \cite{MahantaASS2014} considered the bulk viscous matter to obtain  expression ``(24)". It is to be noted that $\bar P$ in Eqs. ``(22)--(24)" in Mahanta's paper is just a symbol, $P$ with an overhead bar. One may readily verify that there is no use of Eq. ``(14)" to calculate the expression ``(26)" in his work. Hence, for viscous or non-viscous model, one gets the same expressions for the energy density and pressure. Thus, the energy density and pressure obtained in (24) remain independent of bulk viscosity. We shall consider the bulk viscous model in Sect. 3.2.

\subsection{The behavior of coupled matter}

As elucidated above, $\rho_{m}$ and $p_{m}$ do not represent the effective matter in this model of $f(R,T)$ gravity. The terms containing $\lambda$ in Eqs. (21)--(23) can be assumed associated to the coupled matter. By separating these terms, the equations can be expressed as
\begin{eqnarray}
\left(\frac{\dot B}{B}\right)^2+2\frac{\dot A\dot B}{A B}&=&\rho_{m}+\rho_f\\
\left(\frac{\dot B}{B}\right)^2+2\frac{\ddot B}{ B}&=&-(p_m +p_f)\\
\frac{\ddot A}{A}+\frac{\ddot B}{B}+\frac{\dot A \dot B}{AB}&=&-(p_m +p_f),
\end{eqnarray}
where $p_f =\lambda\left(3\rho_m-p_m\right)$ and $p_f =\lambda\left(3p_m -\rho_m\right)$, respectively, represent the energy density and pressure of the coupled matter, and are obtained as
\begin{equation}
\rho_f=p_f=\frac{2\lambda(2n+1)}{(1+2\lambda)(n+2)^2 t^2}.
\end{equation}
Hence, the coupling terms contributes as stiff matter. The energy density and pressure decrease with the evolution. For a physically viable model, the energy density must be positive which is corroborated under the constraints
\begin{eqnarray}\nonumber
-\frac{1}{2}<\lambda <0;\;\;&\text{if}&\;\;n<-\frac{1}{2},\\
\text{or}\;\;\;\lambda <-\frac{1}{2}\;\;\text{or}\;\; \lambda >0; \;\;&\text{if}&\;\;n>-\frac{1}{2}.
\end{eqnarray}
These constraints, in view of (25), agree with $\lambda>0$ and $n>-1/2$ only. Thus, in general, $f(R,T)$ gravity makes the model physically viable for $n<-1/2$ when $\lambda<-1/2$, but if we treat the matter-geometry coupling terms as matter, then the model becomes physically viable only for $\lambda>0$ and $n>-1/2$.

\subsection{Bulk viscous model}

The gravitational field equations with  bulk viscous matter remain the same as given in (21)--(23) or (26)--(28), except that the pressure, $p_m$ is replaced with
\begin{equation}
\bar p_m=p'_m-\xi\theta.
\end{equation}
Now we shall repeat the same procedure that we have followed in Sect. 2.1. First, to examine the behavior of the bulk viscosity coefficient,  we consider the viscous free matter to follow the prefect fluid EoS. Second, by considering the relations of bulk viscosity assumed in cases (i) and (ii) of Sect. 2.1.2, we shall study the behavior of normal matter.

\subsubsection{The behavior of bulk viscous coefficient}

Using the prefect fluid EoS $p'_m=\omega\rho_m$,  where $0\leq\omega\leq1$, we obtain
\begin{equation}
\xi(t) = \frac{(1+2n)(\omega-1)}{(1+2\lambda)\left(2+n\right)^2 t}.
\end{equation}
Since $n>-1/2$ and $\lambda>0$ for a physically viable model, with any kind of matter except stiff fluid, $\xi$ remains negative which increases with the evolution and vanishes at late times. For stiff matter ($\omega=1$), the bulk viscosity coefficient vanishes, and the solutions reduce to the non-viscous model as discussed above. Therefore, the behavior of the bulk viscosity coefficient is similar to the model in GR. Hence, $f(R,T)=R+2\lambda T$ gravity plays no significant role, except that a large value of $\lambda$ diminishes the effect of bulk viscosity.

\subsubsection{The behavior of matter}

\subsubsection*{Case (i) When $\xi \theta= k$}

The EoS parameter of matter, $\omega'_m=p'_m/\rho_m$, gives
\begin{equation}
\omega'_m=1+\frac{k(n+2)^2 (1+2\lambda)t^2}{1+2n}.
\end{equation}
In view of the restrictions $n>-1/2$ and $\lambda>0$, the above EoS parameter for $k>0$ represents semi-realistic matter, whereas for $k<0$, it shows a transition from $\omega'_m=1$ to $\omega'_m\to-\infty$ as $t\to0$, which is similar to the model in GR. Hence, this also indicates that $f(R,T)$ gravity plays no  significant role in this model. However, a large value of $\lambda$ makes the growth (when $k>0$) or reduction (when $k<0$) of $\omega'_m$ much faster. At the origin of evolution, $\omega'_m=1$. If $k=0$, the solutions reduce to the model without bulk viscosity.

\subsubsection*{Case (ii) When $\xi \theta= k_1 \rho$}

The EoS in this case takes a constant value
\begin{equation}
\omega'_m=1+k_1,
\end{equation}
which is identical to (18). Hence, there is no role of $f(R,T)$ gravity in this case.

\section{Conclusion}

Mahanta \cite{MahantaASS2014} studied an LRS Bianchi-I model in $f(R, T)$ gravity with bulk viscous matter. The signs in the field equations in his all the three models of $f(R,T)$ are  incorrect. This minor but serious error makes the model studied by him mathematically, and hence physically, invalid. However, the positive aspect is that the wrong signs do not affect the metric potential. Consequently, the geometrical parameters, namely, volume, expansion scalar, Hubble parameter and shear scalar are  mathematically correct. However, the author skipped the physical interpretation of these geometrical parameters. Later on, Shamir \cite{ShamirEPJC2015} also studied some models without bulk viscosity under the same formulation. He has discussed the geometrical behavior of the model. To obtain the solutions, the authors in the both said works have assumed an expansion scalar proportional to the shear scalar, which returns a constant value of the deceleration parameter, $q=2$. Hence, the model can describe only the decelerated expansion of the universe.

In this paper, we have reconsidered the $f(R,T)=R+2\lambda T$ model studied by Mahanta \cite{MahantaASS2014}. Since a comparison of the outcomes in the modified gravity model with the outcomes of the model in GR helps to understand the role of modified gravity. So before considering the $f(R,T)$ gravity model, we have studied viscous and non-viscous models in GR. A part of our work is also an extension of Shamir's work. Shamir has discussed the geometrical behavior, we have not repeated it here. However, we have shown that these parameters are independent of $f(R,T)$ gravity. Also, while the authors in \cite{MahantaASS2014,ShamirEPJC2015} ignored the testing of physical viability of the models, we have obtained the constraints for a physically realistic cosmological scenario. Mahanta \cite{MahantaASS2014} in Sect. ``3" and ``4.1" merely obtained the expressions of the coefficient of bulk viscosity. Extending the work we have also studied the behavior of normal matter for two different forms of bulk viscosity coefficient considered by him in a model $f(R,T)=R+\lambda T^2$.

The model in GR has been found physically viable only for $n>-1/2$. The effective matter behaves as stiff matter irrespective of a viscous or viscous free model. In the viscous model, the bulk viscosity coefficient with perfect fluid (except for stiff matter) is found to be negative and an  increasing function of cosmic time. In the case of stiff matter, the coefficient of bulk viscosity vanishes. In the reverse approach, with the first assumption $\xi\theta=k$ for $k>0$ the matter follows a semi-realistic EoS, while for $k<0$ the EoS of matter exhibits a transition from a stage of stiff matter to phantom. With the second assumption $\xi\theta=k_1\rho$, the EoS of matter becomes constant ($\omega=1+k_1$), which also renders semi-realistic matter for $k_1>0$, whereas for $k_1<0$ the EoS can describe a variety of matter including radiation, dust, quintessence, phantom, and cosmological constant for different choices of $k_1$. If $k=0=k_1$, the solutions reduce to the model without viscosity.

As far as the $f(R,T)$ gravity model is concerned, Shamir \cite{ShamirEPJC2015} has studied the behavior of effective matter only. However, in case of $f(R,T)$ gravity, some extra terms appear on the right hand side of the field equations. These terms can be treated as representing some additional matter due to the coupling between matter and geometry. Therefore, by considering matter and geometry coupling terms as coupled matter, we have examined its behavior. Since the metric potential remains identical to the model in GR,  the effective matter (irrespective of viscous or non-viscous models) acts as stiff matter in $f(R,T)$ gravity also.

In general, the solutions in $f(R,T)$ gravity are physically viable for $\lambda>-1/2$ and $n>-1/2$ or $\lambda<-1/2$ and $n<-1/2$. However, when the coupling terms are treated as matter then a physically viable model is possible only for $\lambda>0$ and $n>-1/2$. The primary matter as well as coupled matter act as stiff matter. Thus, the behavior of the bulk viscous model in $f(R,T)$ gravity is almost similar to the model in GR. The only difference is that $f(R,T)=R+2\lambda T$ gravity for large values of $\lambda$ diminishes the effect of viscous matter.

Many researchers have been explored cosmological models with stiff matter in the forward approach in different contexts (see for example from \cite{BaliKumawatPLB2008,BanerjeeetalJMP1985,BaliSharmaASS2004,AdhavetalIJTP2008} and references therein). While these works utilize simplified assumptions of the EoS of stiff matter to get exact solutions, it is a natural outcome of the present study. The stiff matter cosmological models are interesting in the sense that for such models the speed of light is equal to the speed of sound \cite{ZeldovichSPJETP1962,ZeldovichMNRAS1972}. A realistic example of the distribution of stiff fluid is a polytropic fluid inside a star. The existence of realistic objects in the universe makes the studies of stiff matter models prominent.

It is also worthwhile mentioning here that Mahanta \cite{MahantaASS2014} considered three models of $f(R,T)$, namely, $f(R,T)=R+2\lambda T$, $f(R,T)=\lambda R+\lambda T$ and $f(R,T)=R+\lambda T^2$. The sign in field equations for all three models is incorrect. The first two forms are, in fact, not different as the first one is a particular case of the second. Consequently, both forms would produce similar results. Moreover, the second model is formulated in a way that the coupling terms are treated as a variable cosmological constant, $\Lambda=(\rho-p)/2$. As we have seen, the energy density and pressure of effective matter as well as coupled matter become equal. Resultantly, $\Lambda$ vanishes in such formulation and the solutions reduce to the model in GR. Consequently, even if one considers the correct sign in the field equations, the outcomes would be identical to the model in GR. Therefore, we have not studied this form explicitly.

Finally, we would like to point out that apart from the wrong signs in the field equations, Mahanta \cite{MahantaASS2014} in his model $f(R,T)=R+\lambda T^2$ over determined the solutions. We see that Eqs. ``(58)--(60)" have five unknowns, namely, $H_1$, $H_3$, $\rho$, $P$ and $\xi$. Therefore, only three assumptions would be required to close the system, but the author used four, namely, ``(27)", ``(28)", ``(61 or 65)" along with the EoS $P=\epsilon \rho$. We have not considered this model in the present study for the sake of keeping our paper of mandate length. Shamir \cite{ShamirEPJC2015} has studied this form without bulk viscosity. We shall consider this model with bulk viscosity somewhere else.

\section*{Acknowledgements} This work is based on the research supported wholly/ in part by the National Research Foundation of South Africa (Grant Numbers: 118511). The authors are grateful to an anonymous referee for constructive criticism which led to an improvement in presentation.


\end{document}